\documentclass[aps,pre,twocolumn,groupedaddress]{revtex4-1}

\usepackage{graphicx}
\usepackage{amssymb}
\usepackage{wasysym}

\begin{document}

\title{Competitive dynamics of two erosion patterns around a cylinder}
\author{F. Lachauss\'ee$^1$, Y. Bertho$^1$, C. Morize$^1$, A. Sauret$^{1,2}$ and P. Gondret$^1$}
\affiliation{$^1$ Laboratoire FAST, Univ. Paris-Sud, CNRS, Universit\'e Paris-Saclay, F-91405, Orsay, France}
\affiliation{$^2$ Laboratoire SVI, CNRS, Saint-Gobain, F-93303 Aubervilliers, France}

\begin{abstract}
We investigate experimentally the local erosion of a granular bed near a fixed vertical cylinder that emerges from the bed. The onset of erosion arising at the base of the cylinder and usually ascribed to the wrapping horseshoe vortex is determined and rationalized by a flow contraction effect. We report a new erosion pattern visible downstream of the cylinder that consists of two side-by-side elongated holes. This pattern is observed for flow regimes close to the horseshoe scour onset, whose growth usually inhibits its spatiotemporal development.
\end{abstract}

\maketitle

\section{Introduction}

Granular material is removed and transported by fluid flow through erosion processes, which are encountered in numerous natural and industrial situations. Understanding and modeling the erosion processes are essential to study geomorphological patterns or to predict the evolution of river beds \cite{Huggett2007, Ristroph2012, Andreotti2013}. The erosion begins when a critical shear stress is reached. Defining this onset is, therefore, key to determining the subsequent law for grain transport \cite{Yalin1976, Houssais2017}. Well-controlled laboratory experiments constitute an opportunity to investigate the early times of the erosion \cite{Huppert1986}. Indeed, determining the onset, even under the action of a simple shear flow, remains challenging as it is intimately related to the solid-liquid transition for the granular matter \cite{Loiseleux2005, Ouriemi2007, Houssais2015, Yan2016, Aussillous2016}.

Quantifying the erosion in the vicinity of structures is complex as the fluid flow becomes complex with the presence of vortices. As erosion can reduce the stability and integrity of a wide range of hydraulic structures, its characterization would benefit the design and risk assessment of engineered systems such as bridge piers, offshore platforms, and wind or hydraulic turbines. Past work in large-scale facilities \cite{Breusers1977, Ettema2011} has not allowed the development of robust physical models for the erosion that occurs around structures \cite{Roulund2005, Manes2015} because of both the rheology of granular materials \cite{Andreotti2013} and the complexity of the flow generated around the structure \cite{Williamson1996, Unger2007}. Indeed, in the vicinity of a cylinder, several types of vortices are generated such as the horseshoe vortex that wraps around the base of the cylinder or the B\'enard-von K\'arm\'an vortices downstream of the cylinder. In recent decades, most studies have focused on the time evolution of scouring and on the prediction of the maximum scour depth \cite{Ettema2011,Manes2015}. Surprisingly, only little attention has been devoted to the onset of erosion, which is usually not taken into account in erosion modeling \cite{Manes2015}. However, the threshold of erosion is a key parameter to improve the models of erosion significantly and describe the transport processes.

In this paper, using small-scale laboratory experiments, we investigate the erosion threshold and the resulting patterns around an immersed cylinder. We specifically focus on the clear-water regime, which corresponds to the situation where no erosion occurs in the absence of the obstacle. In addition to the traditional horseshoe scouring, we report a new erosion pattern downstream of the cylinder, consisting of two side-by-side elongated holes. We study on the onset and the competitive growth dynamics of these two erosion patterns.

\section{Experimental methods}

\begin{figure}[b]
\centering
\includegraphics[width=0.8\columnwidth]{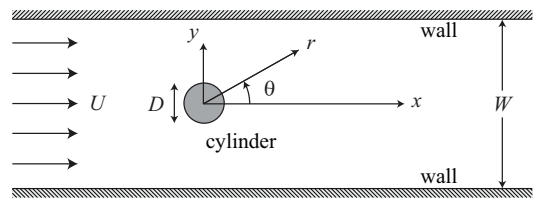}
\caption{Schematic of the experimental setup (top view).}
\label{fig1}
\end{figure}

The experimental setup used consists of a closed racetrack-shaped channel 3.6\,m long and $W=0.1$\,m wide, composed of two linear parts, each of length $L=0.9$\,m and separated by two opposite curved parts with a radius of 0.3\,m. A similar setup was recently used to determine the onset of erosion of a granular bed without obstacle \cite{Salevan2017}. A paddle wheel located in one linear part of the channel generates the fluid flow. In the opposite linear test section, a granular bed made of sieved glass beads of density $\rho_g=2.5 \times 10^3$\,kg\,m$^{-3}$ and diameter $d=0.27\pm 0.03$\,mm is put in a removable drawer of 600\,mm length and 40\,mm depth at the bottom of the channel. A vertical cylinder of diameter $D$ in the range $5 \leqslant D\leqslant 30$\,mm is fixed on the granular bed at a mid-gap width and $520$\,mm away from the entrance of the linear test section. The water height used in the experiments remains constant and equal to $h=160$\,mm. The experiments have been performed with a cylinder of height $H=90$\,mm above the granular bed. The side walls of the test section are made of glass and allow a direct visualization of the scouring in the vicinity of the cylinder. A schematic of the setup and the coordinate system is shown in Fig.~\ref{fig1}. A laser profilometer is mounted on a horizontal translation guide to scan the granular bed topography from above, every 5\,min without stopping the flow. We denote by $\xi(x,y,t)$ the topography corresponding to the difference from the initial horizontal topography measured before any flow at $t < 0$. We also put a horizontal glass plate on the water surface above the granular bed to limit the signal noise in topography measurements that may arise from the fluctuations of the water surface.

Before each experiment, the granular bed is stirred and leveled with a ruler, which ensures a reproducible initial packing and a flat horizontal surface. The upper part of the cylinder is then magnetically fixed on its lower part previously buried in the granular medium and flushed with the granular bed. The flow is then switched on at a given velocity $U$ at time $t=0$. The flow velocity is controlled by the rotation of the paddle wheel, and is up to a mean value of the flow $U\simeq 0.25$\,m\,s$^{-1}$. The typical Reynolds number based on the channel width $W$ is $\mathrm{Re}_W =\rho UW/\eta \simeq 10^4$ considering the density $\rho =10^3$\,kg\,m$^{-3}$ and viscosity $\eta =10^{-3}$\,Pa\,s of water. The channel flow is thus turbulent, and the velocity profiles are measured using particle image velocimetry (PIV) methods. We ensure that the velocity profile just before the cylinder is nearly the same throughout the channel width, with an adequate honeycomb filter at the entrance of the linear test section. The velocity profile has a nearly flat profile except near the bottom surface with a friction velocity $u^* \simeq U/13$, corresponding to the boundary layer thickness $\delta \simeq 1$\,mm. The Reynolds number of the flow around the cylinder is $\mathrm{Re}_D\simeq 10^3$, far beyond the critical values ranging from 50 to 90 for the B\'enard--von K\'arm\'an unsteadiness for an infinite cylinder between two parallel sidewalls with the present aspect ratio $0.05\leq D/W\leq 0.3$ \cite{Chen1995}. The flow around the cylinder is thus also turbulent with strong unsteady wake vortices downstream. Finally, although our experiments are performed with a horizontal glass plate at the free surface, the results presented here are also valid for a free-surface flow. Indeed, the Froude number $\mathrm{Fr}=U/(gh)^{1/2}$, which describes the ratio of the flow velocity to the velocity of surface waves, is smaller than 0.2 and thus the interaction of the fluid free surface with the deformation of the granular bed can be neglected.

\section{Erosion patterns}

Experimentally, the erosion of the granular bed without any cylinder occurs above the threshold velocity $U_c \simeq 0.17$\,m\,s$^{-1}$, corresponding to a global Shields number $\mathrm{Sh}_c = \rho U^2/(\Delta\rho gd) \simeq 7.5$. The global Shields number represents the ratio of the destabilizing global inertial fluid stress $\rho U^2$ far from the bed surface to the stabilizing reduced gravity stress on the grains $\Delta\rho \, g\,d$, where $\Delta\rho = \rho_g - \rho$ is the density difference between the grains and the fluid. Here, we consider the inertial Shields number as the particulate Reynolds number $\mathrm{Re}_d\simeq 50$ is larger than unity, which corresponds to an inertial regime. The local Shields number, based on the fluid stress estimated at the bed surface, is $\mathrm{Sh}_c^* \simeq \mathrm{Sh}_c/170 \simeq 0.04$ at the onset of erosion. This value is in the range of threshold values usually reported for the present shear Reynolds number $\mathrm{Re}^* \simeq 3$ based on the friction velocity $u^*$ and grain diameter $d$ \cite{Clark2017}. We should emphasize that different methods are used to determine the onset of erosion \cite{Houssais2015} and we choose to measure the onset visually by increasing the flow velocity slowly by steps of 1\,mm\,s$^{-1}$ until a few grains are observed to move, over a few minutes. This procedure is repeated to quantify the dispersion of the measurements around the mean value. In the following, we will consider the influence of the cylinder, and characterize the erosive strength of the flow with the reduced Shields number $\mathrm{Sh}/\mathrm{Sh}_c$ defined as the ratio of the Shields number to the value $\mathrm{Sh}_c$ obtained for the live-bed condition, \textit{i.e.} when the sediment transport occurs upstream of the obstacle. The value $\mathrm{Sh}/\mathrm{Sh}_c = 1$ thus corresponds to the live-bed condition whereas $\mathrm{Sh}/\mathrm{Sh}_c = 0$ corresponds to no fluid flow.

\begin{figure*}
\centering
\includegraphics[width=2\columnwidth]{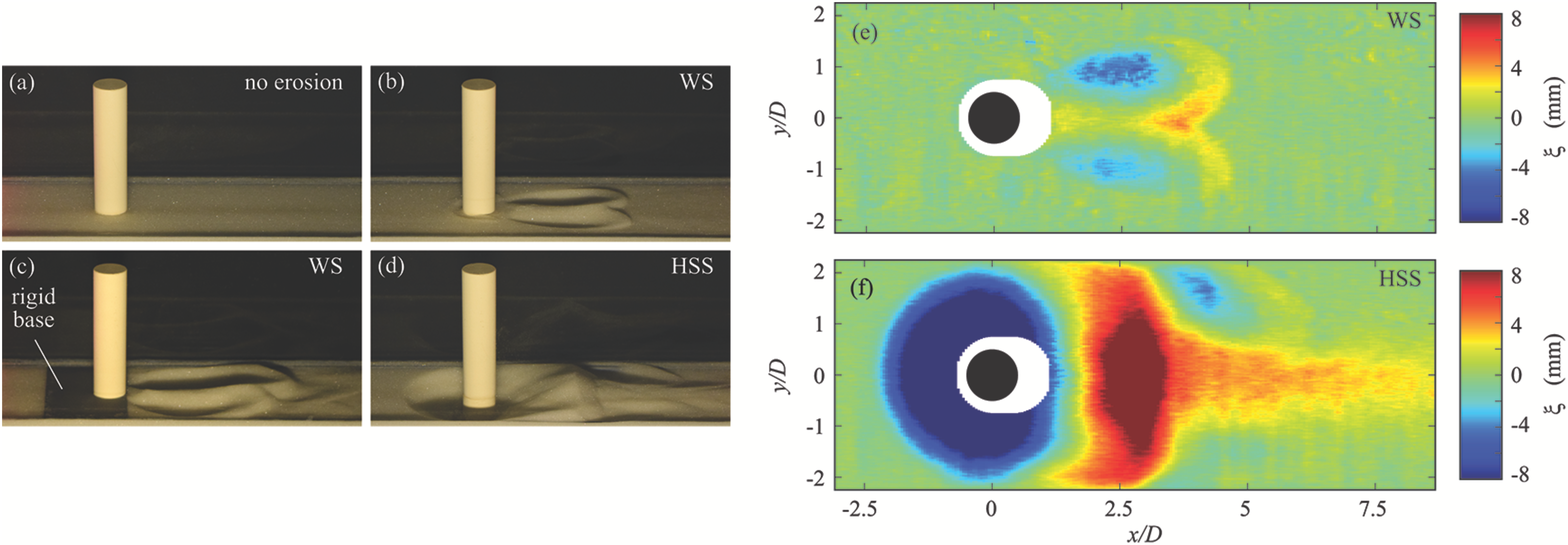}
\caption{Pictures of the granular bed in the vicinity of the cylinder of diameter $D=20$\,mm for (a)~$\mathrm{Sh}/\mathrm{Sh}_c \simeq 0.10$, $\mathrm{Re}_D\simeq 1000$ and $t\simeq 3.0$\,h, (b)~$\mathrm{Sh}/\mathrm{Sh}_c\simeq 0.37$, $\mathrm{Re}_D\simeq 2100$ and $t\simeq 6.9$\,h, (c)~$\mathrm{Sh}/\mathrm{Sh}_c \simeq 0.85$, $\mathrm{Re}_D\simeq 3200$ and $t\simeq 0.72$\,h, with a rigid base at the bottom, and (d)~$\mathrm{Sh}/\mathrm{Sh}_c \simeq 0.85$, $\mathrm{Re}_D\simeq 3200$ and $t\simeq 0.17$\,h. Also shown is the topography $\xi(x,y$) of the (e)~wake scour (WS) and (f)~horseshoe scour (HSS) patterns for the same parameters as in (b) and (d), respectively.}
\label{fig2}
\end{figure*}

\medskip

We now consider the scouring in the presence of the cylinder in clear-water conditions, for which $\mathrm{Sh}/\mathrm{Sh}_c<1$. Erosion takes place in the vicinity of the cylinder where the flow is modified by the immersed obstacle. Figures~\ref{fig2}(a)-\ref{fig2}(d) show the scour pattern, imaged from the side, which occurs around a cylinder of diameter $D=20$\,mm for various values of $\mathrm{Sh}$ and $\mathrm{Re}_D$. At low enough $\mathrm{Sh}$, no erosion occurs, even in the presence of an obstacle [Fig.~\ref{fig2}(a)]. At large enough $\mathrm{Sh}$ values, a large hole with an annular shape is observed all around the cylinder as shown in Fig.~\ref{fig2}(d). The corresponding topography measurements reported in Fig.~\ref{fig2}(f) exhibit a large hole all around the cylinder with a nearly circular shape of diameter $\simeq 4D$ and maximal depth $\simeq 0.4D$, flattened at the rear. We also observe a dune formed by the scoured grains downstream with a roughly equivalent maximal height $\simeq 0.4D$. Here, the Shields number is $\mathrm{Sh} \simeq 6.4$, close to the value measured in the live-bed condition, $\mathrm{Sh}_c\simeq 7.5$. This value corresponds to the reduced Shields number $\mathrm{Sh}/\mathrm{Sh}_c \simeq 0.85$, so that the scouring phenomenon is almost maximal. This erosion pattern localized at the base of the cylinder has been widely observed and is usually attributed to the erosive action of the horseshoe vortex that forms just upstream of the cylinder near the bottom surface and wraps around the cylinder as a necklace \cite{Breusers1977, Ettema2011, Roulund2005, Manes2015}. We will thus refer to such an erosion process as ``horseshoe scour'' (HSS). For smaller flow values, the erosion is weaker and vanishes for the critical Shields value $\mathrm{Sh}_c^\mathrm{HSS} \simeq 2.1$ corresponding to $\mathrm{Sh}_c^\mathrm{HSS}/\mathrm{Sh}_c \simeq 0.28$ in the case of a cylinder of diameter $D=20$\,mm.

\medskip

We also performed long experiments close to the HSS onset. In this condition, we observed a new erosion pattern presented in Fig.~\ref{fig2}(b). This pattern is observed downstream of the cylinder with two side-by-side elongated holes, as shown by the topography measurements in Fig.~\ref{fig2}(e). This pattern is called ``wake scour'' (WS) pattern in the following. In Fig.~\ref{fig2}(b), the Shields number is $\mathrm{Sh} \simeq 2.8$ corresponding to the reduced Shields number $\mathrm{Sh}/\mathrm{Sh}_c \simeq 0.37$, just a little above the HSS onset, measured at $\mathrm{Sh}_c^\mathrm{HSS}/\mathrm{Sh}_c \simeq 0.28$. This explains why a weak HSS close to the cylinder is also observed here. Wake scour is here stronger than HSS, with two ovoid holes, symmetrical with respect to the $x$-axis, and measuring about $2\,D$ in length and $1\,D$ in width, and located at $(x,y) \simeq (2\,D,\pm 1D)$. For larger flow values, the development of the HSS pattern is quicker so that it progressively impedes the WS development. We should note that a partial WS pattern can be observed in Fig.~\ref{fig2}(d), with two lateral elongated holes downstream of the HSS at $(x,y) \simeq (4D,\pm 1.5D)$. For smaller flow values, WS vanishes for the critical value $\mathrm{Sh}_c^\mathrm{WS} \simeq 0.9$ corresponding to $\mathrm{Sh}_c^\mathrm{WS}/\mathrm{Sh}_c \simeq 0.12$ in the case of a cylinder of diameter $D = 20$\,mm.

\begin{figure}[b]
\centering
\includegraphics[width=0.75\columnwidth]{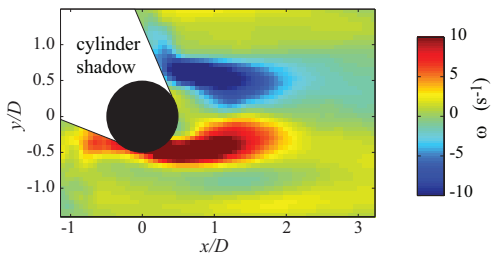}
\caption{Average vorticity field $\omega$ downstream of a cylinder of diameter $D=20$\,mm. Particle image velocimetry measurements are performed in a horizontal plane located 5\,mm above a rough non-erodible bed, for $\mathrm{Sh}/\mathrm{Sh}_c \simeq 0.22$ and $\mathrm{Re}_D\simeq 1600$.}
\label{fig3}
\end{figure}

For a rather large value of $\mathrm{Sh}$ such as the case presented in Fig.~\ref{fig2}(d) ($\mathrm{Sh}/\mathrm{Sh}_c \simeq 0.85$), the HSS development is not affected by WS, but the WS development is in contrast strongly affected by HSS. Therefore, to characterize the WS independently of HSS, we put a rigid base at the bottom of the cylinder in a rectangular zone of width $W$ and length $2D$ for $-3D/2 \leqslant x \leqslant D/2$ as shown in Fig.~\ref{fig2}(c). In that case, the HSS development is inhibited as the granular bed cannot be eroded from streamlines contraction and horseshoe vortex close to the cylinder. By visual observation of the grain motion and from PIV measurements of the flow in several horizontal planes downstream of the cylinder, we believe that the WS pattern arises from the unsteady B\'enard--von K\'arm\'an lee vortices. The WS pattern first develops just downstream of the cylinder, where the vorticity is maximal as illustrated in Fig.~\ref{fig3}, before extending into two elongated holes. We performed PIV measurements 5 mm above a rough non-erodible bed so that the velocity field corresponds to the one existing before any bed deformation at the beginning of an erosion experiment. The regions of large vorticity correspond to the vortex shedding area, where grains are carried by vortices and advected downstream, thus initializing the formation of the WS pattern. At high Reynolds numbers, far from the B\'enard--von K\'arm\'an onset, the maximum vorticity is observed at a distance of about $5D$ downstream if an infinite cylinder \cite{Goujon1994}. However, in the present case where the cylinder emerges from the granular bed, the maximum vorticity occurs at a smaller distance. Similar erosion patterns have already been observed downstream of a cylinder but were attributed to a secondary effect of the HSS development far above the HSS onset \cite{Auzerais2016}. However, the WS pattern reported here has been shown to exist without any HSS development and is thus not linked to HSS.

\section{Erosion thresholds and dynamics}

\begin{figure}[t]
\centering
\includegraphics[width=0.75\columnwidth]{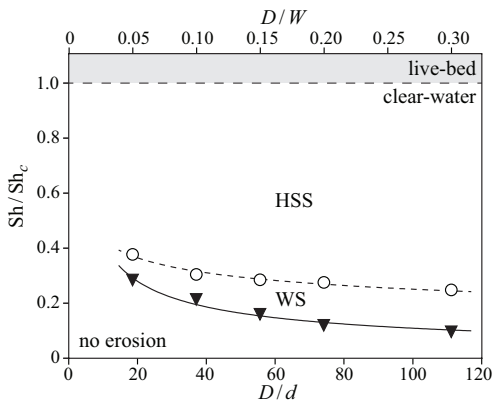}
\caption{Critical values of the reduced Shields number $\mathrm{Sh}/\mathrm{Sh}_c$ for the scour onset as a function of the cylinder to grain size ratio $D/d$ and cylinder to channel size ratio $D/W$ for the two observed patterns: HSS ($\circ$) and WS ($\blacktriangledown$). The horizontal dashed line $\mathrm{Sh}/\mathrm{Sh}_c =1$ corresponds to the onset of erosion without any cylinder. The curved dashed and solid lines correspond to power laws $\mathrm{Sh}/\mathrm{Sh}_c = \alpha (D/d)^\beta$ with the best fitting values $\alpha = 0.73$ and $\beta = -0.23$ for HSS (-~-~-), and $\alpha = 1.60$ and $\beta = -0.58 \pm 0.1$ for WS (---).}
\label{fig4}
\end{figure}

Onset measurements for HSS are reported in Fig.~\ref{fig4} for different cylinder diameters in the range $5 \leq D \leq 30$\,mm. The erosion threshold for HSS, $\mathrm{Sh}_c^\mathrm{HSS}$, does not depend much on $D$ in this range. Yet, a gradual increase is observed from the value $\mathrm{Sh}_c^\mathrm{HSS}/\mathrm{Sh}_c \simeq 0.25$ for the largest cylinder to $\mathrm{Sh}_c^\mathrm{HSS}/\mathrm{Sh}_c \simeq 0.38$ for the smallest one. Onset measurements for WS, $\mathrm{Sh}_c^\mathrm{WS}$, are also reported and follow a similar trend to the HSS for all the cylinder diameters considered, with smaller values from about 0.10 to 0.29. Below $\mathrm{Sh}_c^\mathrm{WS}$, no erosion occurs on the bed surface around or far from the cylinder. A power law $\mathrm{Sh}/\mathrm{Sh}_c = \alpha (D/d)^\beta$ passes rather well through each data set with the fitting values $\alpha = 0.73$ and $\beta = -0.23$ for HSS, and $\alpha = 1.60$ and $\beta = -0.58$ for WS. The cylinder size has thus a stronger influence on WS onset than on HS onset.

To rationalize the location and the onset of the HSS pattern, we observe that erosion does not occur first upstream of the cylinder at $(r,\theta) \simeq (D/2,\,\pi)$, but at the cylinder sides $(r,\theta) \simeq (D/2,\pm 7\,\pi/10)$. This location corresponds to the region where the shear stress exerted on the granular bed is the largest, due to both the streamlines contraction and horseshoe vortex \cite{Roulund2005}. A two-dimensional potential flow theory around an infinite cylinder can capture the streamlines contraction effect. In an unbounded domain, the stream function is $\psi = Ur \sin \theta\left [ 1-(D/2r)^2\right ]$ and leads to a maximal velocity $u_\mathrm{max}=2U$, corresponding to the azimuthal velocity $u_\theta = -\partial \psi/ \partial r$ at $(r,\theta) = (D/2,\pm \pi/2)$ \cite{Guyon2015}. Since the local erosion takes place at the critical velocity $U_c$, the incoming velocity far from the cylinder that corresponds to the HSS onset would be $U_c^\mathrm{HSS}=U_c/2$. Therefore, the threshold value of the global Shields number would be $\mathrm{Sh}_c^\mathrm{HSS} = \mathrm{Sh}_c/4$. The critical value $\mathrm{Sh}_c^\mathrm{HSS}/\mathrm{Sh}_c = 0.25$ independent of the cylinder diameter given by this very simplified potential flow approach is close to the experimental measurements.
Let us now explore some possible effect of the channel width by still considering a two-dimensional potential flow around an infinite cylinder but now at mid-gap between two lateral walls at the distance $W$. Following Lamb's approach \cite{Lamb1932} with a uniform flow across a periodic line of dipoles at the distance $W$ from one another which leads to an analogous perturbed flow, the maximal flow velocity $u_\mathrm{max}$ still occurs at $(r,\theta) = (D/2,\pm \pi/2)$ but depends on the blockage ratio $D/W$ as
\begin{equation}
u_\mathrm{max} = U \left[1 + \pi \frac{D}{W} \frac{\tan\left(\pi D/2 W\right)}{1 - \cos\left(\pi D/W\right)} \right].
\end{equation}
For vanishing blockage ratio ($D/W \rightarrow 0$), we recover the value $u_\mathrm{max}=2U$, whereas for our largest blockage ratio $D/W = 0.3$, this modeling leads to a slightly larger velocity $u_{max}=2.16U$, which corresponds to the critical reduced Shields number parameter $\mathrm{Sh}_c^\mathrm{HSS}/\mathrm{Sh}_c = 0.21$ slightly smaller than the value 0.25 predicted for vanishing $D/W$. Note that this modeling leads to some non-circularity of the cylinder for a very large blockage ratio. For our largest ratio $D/W = 0.3$, the non-circularity is only 0.2\%. From the present analysis of the finite size of the channel, the variation of $\mathrm{Sh}_c^\mathrm{HSS}$ is thus expected to be of only 15\%, which is not sufficient to explain the 35\% variation measured experimentally. A little more refined two-dimensional theory taking into account the boundary layer development at the cylinder surface would not lead to a significantly different prediction, and a three-dimensional theory taking into account the boundary layer on the bottom plate and the horseshoe vortex is much more complex. However, the present simple two-dimensional potential flow catches the good first order of the HSS onset, which means that HSS arises mainly from the streamlines contraction.

\begin{figure}
\centering
\includegraphics[width=0.75\columnwidth]{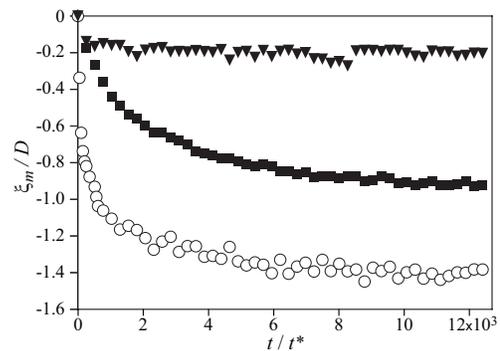}
\caption{Dimensionless scour depth $\xi_m/D$ close to a cylinder of diameter $D=10$\,mm as a function of the dimensionless time $t/t^* = t (\rho d/\Delta \rho g)^{-1/2}(D/d)^{-2}$, for the HSS pattern at $\mathrm{Sh}/\mathrm{Sh}_c \simeq 0.69$~($\circ$), and the WS pattern at $\mathrm{Sh}/\mathrm{Sh}_c \simeq 0.40$~($\blacktriangledown$) and $\mathrm{Sh}/\mathrm{Sh}_c \simeq 0.69$~($\blacksquare$) by using a rigid bed at the bottom of the cylinder.}
\label{fig5}
\end{figure}

From the time evolution $\xi(x,y,t)$ of the bed surface reported in Figs.~\ref{fig2}(e) and \ref{fig2}(f) for the WS and HSS, respectively, we extract the time evolution of the maximum scour depth $\xi_m(t)$. The results are reported in Fig.~\ref{fig5} for WS pattern at $\mathrm{Sh}/\mathrm{Sh}_c \simeq 0.40$ and for HSS and WS patterns at the same erosion parameter $\mathrm{Sh}/\mathrm{Sh}_c \simeq 0.69$. For both erosion patterns, the scouring depth $\xi_m$ increases with time and seems to saturate at a plateau value $\xi_\infty$ at large times. In Fig.~\ref{fig5}, scouring depth $\xi_m$ and time $t$ are made dimensionless with the cylinder diameter $D$ and with the typical time scale $t^*=(\rho d/\Delta \rho g)^{1/2}(D/d)^{2} \simeq 6$\,s, corresponding to an inertial (turbulent) grain transport over the typical eroded area $D^2$ \cite{Andreotti2013}. The long-time depth of the two scouring patterns obtained here at a similar value $\mathrm{Sh}/\mathrm{Sh}_c \simeq 0.69$ is $\xi_\infty \simeq -1.4D$ for HSS and $\xi_\infty \simeq -0.8D$ for WS. The characteristic time of evolution of the two scouring patterns is $\tau \simeq 0.35$\,h $\simeq 210 t^*$ for HSS and $\tau \simeq 2$\,h $\simeq 1220 t^*$ for WS. Therefore, at $\mathrm{Sh}/\mathrm{Sh}_c \simeq 0.69$, far above both the WS and HSS onsets, the HSS development occurs faster and over a larger depth. Since the time evolution of the maximal scour depth cannot be fitted by a simple law, $\tau$ is here defined as the time necessary to reach half of the plateau value.

\begin{figure}
\centering
\includegraphics[width=\columnwidth]{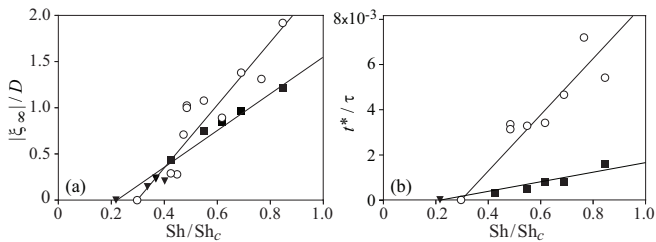}
\caption{Evolution of (a)~the dimensionless long-time scour depth $| \xi_\infty |/D$ and (b)~the inverse of the dimensionless characteristic time evolution of scouring $t^*/\tau$ in the vicinity of a cylinder of diameter $D=10$\,mm as a function of the reduced Shields number $\mathrm{Sh}/\mathrm{Sh}_c$ for the HSS pattern~($\circ$) and the WS pattern without~($\blacktriangledown$) or with~($\blacksquare$) a non-erodible base at the bottom of the cylinder.}
\label{fig6}
\end{figure}

The dimensionless depth at long time $\xi_\infty/D$ and the dimensionless characteristic time of scouring $\tau/t^*$ are shown in Fig.~\ref{fig6} as a function of the reduced Shields number $\mathrm{Sh}/\mathrm{Sh}_c$. For decreasing $\mathrm{Sh}$, we observe that the efficiency of scouring processes decreases, so $|\xi_\infty |$ decreases and $\tau$ increases. At the threshold, $|\xi_\infty|$ vanishes whereas $\tau$ diverges, so $1/\tau$ also vanishes. Despite the dispersion of the experimental data, $|\xi_\infty|$ and $1/\tau$ may be approximated by a linear increase of $\mathrm{Sh}$ above the corresponding critical value at onset $\mathrm{Sh}_c^\mathrm{HSS}$ or $\mathrm{Sh}_c^\mathrm{WS}$ measured previously (Fig.~\ref{fig4}). With the linear fits shown in Fig.~\ref{fig6}, we see that WS is stronger than HSS for $0.2 \lesssim \mathrm{Sh} \lesssim 0.4$ thus in a small region compared to the large zone $0.4 \lesssim \mathrm{Sh} \leqslant 1$ where HSS is stronger than WS.

By an interesting scaling analysis of the complex flow around vertical cylinder interacting with a horizontal granular bed within the turbulent regime, Manes and Brocchini~\cite{Manes2015} have derived the simple scaling $|\xi_\infty |/d \sim \mathrm{Sh} (D/d)^{2/3}$ for the HSS pattern. Such a scaling is consistent with the cloud of data they collected from the literature with large hydraulic channel facilities. Our own $\xi_\infty$ experimental data are within this cloud of collected data. We believe, however, that some of the dispersion of the collected data relative to the theoretical scaling proposed in \cite{Manes2015} arises from the fact that no erosion threshold is taken into account in this scaling. As a matter of fact, we see that our $\xi_\infty$ data scales about linearly with $\mathrm{Sh}-\mathrm{Sh}_c^\mathrm{HSS}$ but not with $\mathrm{Sh}$. Note that the theoretical scaling of \cite{Manes2015} corresponds to $|\xi_\infty/D| \sim \mathrm{Sh} (D/d)^{-1/3}$ with thus a weak dependence on $D/d$ in the range $D/d \gg 1$ where this scaling is valid. It is worth noting that our measurements for the critical Shields number for HSS onset present also only a weak dependence on $D/d$ with a power exponent -0.23 not very far from -1/3.

\section{Conclusion}
In summary, using a laboratory setup, we highlighted in this paper the competitive dynamics of two erosion patterns in the vicinity of a vertical cylinder emerging from a granular bed when submitted to a steady flow in clear-water conditions. Our experiments revealed that the onset of the usual HorseShoe Scour (HSS) pattern, which arises at the bottom of the cylinder, does not depend significantly on the cylinder diameter and the threshold value $\mathrm{Sh}_c^\mathrm{HSS}$ can be rationalized considering the streamline contraction inducing a more intense erosive strength. Slightly below the onset for HSS, we measured the onset of another erosion pattern designed as WS, which appears downstream of the cylinder and exhibits a peculiar geometry, two ovoid holes. This new pattern reported here is related to the local erosive action of the downstream unsteady lee-vortices in the cylinder wake. Wake scour is stronger than HSS in only a very small region of flow conditions, which could explain why it has not been reported yet. The three-dimensional flow around the cylinder and the complex granular rheology make a quantitative modeling and numerical simulations challenging \cite{Roulund2005}.

\section*{Acknowledgements}
This work was supported by a public grant from the French National Research Agency within the project SSHEAR (Grant No. ANR-14-CE03-0011). We thank J. Amarni, A. Aubertin, L. Auffray, and R. Pidoux for their contribution to the development of the experimental setup.

\bibliography{biblio}

\end{document}